\documentstyle[sprocl]{article}

\def\Journal#1#2#3#4{{#1} {\bf #2}, #3 (#4)}

\def\JCP{\em J. Chem. Phys.} 
\def\IJMP{{\em Int. J. Mod. Phys.} E}

\def\NPB{{\em Nucl. Phys.} B}

\def\PLA{{\em Phys. Lett.}  A}
\def\PRL{\em Phys. Rev. Lett.}
\def\PRE{{\em Phys. Rev.} E}
\def\PRD{{\em Phys. Rev.} D}

\def\be{\begin{equation}}
\def\ee{\end{equation}}
\def\bea{\begin{eqnarray}}
\def\eea{\end{eqnarray}}

\begin{document}

\title{UNVERSAL FEATURES OF THE ORDER-PARAMETER FLUCTUATIONS}

\author{R. BOTET}

\address{LPS, Universit\'{e} Paris-Sud, Centre d'Orsay, \\F-91405 Orsay,
France\\E-mail: botet@lps.u-psud.fr}

\author{M. PLOSZAJCZAK}

\address{Grand Acc\'{e}l\'{e}rateir National d'Ions Lourds, BP 5027,\\ 
F-14076 Caen, France \\E-MAIL: ploszajczak@ganac4.in2p3.fr}         


\maketitle\abstracts{We discuss the universal scaling laws of order parameter
fluctuations in any system in which the second-order critical behavior can be
identified. These scaling laws can be derived rigorously 
for equilibrium systems 
when combined with the finite-size scaling analysis. The relation 
between order
parameter, criticality and scaling law of fluctuations has been established
and the connexion between the scaling function
and the critical exponents has been found.}
 
\section{Introduction}\label{sec:intro} 
Averaged quantities are still commonly used 
for the description of many complex processes in physics, chemistry or
astrophysics. This kind of approach is inprinted in our minds by an
unquestionable success of thermodynamics in which fluctuations around the
thermodynamic value are small in large systems. It is our experience that the
gross measures in, {\it e.g.}, the particle reaction data are not
sufficient to discriminate among models unless supplemented with more fine
grained information, especially fluctuations and correlations of various kinds.
Recent resurgence of interest in fluctuations in the 
strong interaction physics is due to
the experimental possibility to measure event by event 
many strongly fluctuating quantities, {\it e.g.}, the multiplicity
of produced particles in ultrarelativistic collisions of leptons, hadrons,
nuclei, or the charges (masses) of 
fragmentation products of highly excited heavy-ion residue.
The immediate question is then what is the 
information contained in fluctuations and, in particular, 
in the order parameter fluctuations
, how do these fluctuations scale with the
system-size and what is the relation between criticality and the probability
distribution of order parameter fluctuations.

\section{Order parameter fluctuations}\label{sec:order} 
Let us suppose that the thermodynamic free energy of finite system 
depends on three
parameters : $\eta$ (the intensive order parameter), $\epsilon$~ (the distance
to the critical point) and $N$ (the size of the system).
Widom \cite{Widom}~ has proposed that close to the critical point,
the free energy density in the thermodynamic limit scales as :
\begin{eqnarray}
\label{wid1}
f_o (\lambda^{\beta} \eta, \lambda \epsilon) \sim 
\lambda^{2-\alpha} f_o (\eta, \epsilon) ~~~ \ ,
\end{eqnarray}
where $\alpha, \beta $ are usual critical exponents. 
There is no critical behavior in a finite system. 
However, a finite system behaves like an infinite one 
if the correlation length $\xi $  becomes comparable to 
the typical length $L$~ of the system. This is basically the
argument of Fisher and Barber \cite{FisherBarber}~ 
leading to  the finite-size scaling
analysis of critical systems. The pseudocritical point for a
finite system appears at a distance  $\epsilon \sim 
c N^{-1/{\nu d}}$~ 
from the critical point\cite{FF}~, where $c$~ is some 
dimensionless constant
which can be either positive or negative\cite{comment}.
One can then deduce the scaling of 
critical free energy density at this point :
\begin{eqnarray}
\label{Widomfinite}
f_c(\eta ,N) \sim \eta^{\frac{2-\alpha }{\beta }} 
\phi(\eta 
N^{\frac{\beta }{\nu d}})~~~ \ .
\end{eqnarray}
Assuming the hyperscaling relation : $2-\alpha =  \nu 
d$~, and 
using Rushbrooke relation between critical exponents  
: $\alpha + 2 \beta + \gamma = 2$, one can write
the total free energy $F(\eta , \epsilon, N) = N f_o 
(\eta , \epsilon )$~ 
at the pseudocritical point as follows :
\begin{eqnarray}
\label{psi}
F_c(\eta , N) \sim f_o (\eta N^{\frac{\beta }{\gamma + 2 
\beta }},c)~~~ \ .
\end{eqnarray}
The canonical probability density $P[\eta ]$ to get some 
value of the order parameter $\eta $~ is given by\cite{Mayer}~:
\begin{eqnarray}
\label{proz}
P[\eta ] = {Z_N}^{-1} \exp(-{\beta}_T F(\eta , \epsilon , N)) ~~~ \ ,
\end{eqnarray}
where ${\beta}_T$~ is the inverse of temperature. Using Eq.
(\ref{psi}), one obtains the partition function : 
\begin{eqnarray}
\label{part}
Z_N \sim N^{-\frac{\beta }{\gamma + 2 \beta}} \sim 
<|\eta |> ~~~\ . 
\end{eqnarray}
It is then easy to see that the probability density 
$P[\eta ]$~ obeys the first scaling law :
\begin{eqnarray}
\label{first}
<|\eta |> P[\eta ] & = & \Phi (z_{(1)}) =  \Phi \left( 
\frac{\eta - <| \eta |>}{<| \eta |>} \right) \nonumber \\
& = & a({\beta}_T ) \exp \left( - {\beta}_T 
f_o \left( \frac{\eta}{ <| \eta |>}, c \right) \right) 
~~~ \ ,
\end{eqnarray}
with the constant coefficient ${\beta}_T$~ independent of $\eta $~, and
the scaling function $\Phi (z_{(1)})$~ depending on a single 
scaled variable : $z_{(1)} = (\eta - <| \eta |>)/<| \eta 
| >$~. The scaling limit is defined by the asymptotic behaviour of $P[ \eta ]$
when $\eta \rightarrow \infty$, $<| \eta |> \rightarrow \infty$, but 
$({\eta}/{<| \eta |>})$~ has a finite value.
The temperature-dependent factor $a({\beta}_T)$  is determined by the
normalization of $P[\eta ]$~. One may notice that the logarithm of 
scaling function (\ref{first}) corresponds 
to the non-critical free energy density
at the renormalized distance $\epsilon =c$~ from the 
critical point. If the
order parameter corresponds to the cluster multiplicity, like in the
fragmentation - inactivation binary (FIB) 
process\cite{sing1,sing11}, then (\ref{first})
can be written in an equivalent form to the KNO 
scaling\cite{koba}~,
proposed some time ago as the ultimate symmetry of $S$ - matrix in 
the relativistic field theory\cite{comment1}~.
Relation between the KNO scaling and the phase 
transition in Feynman-Wilson gas as well as the criticality of self-similar
FIB process was studied as well\cite{antoniou,kno}~.
Defining the anomalous dimension for an extensive quantity $N \eta$~ as : 
\begin{eqnarray}
\label{anomdim}
g = {\lim}_{N \rightarrow \infty} g_N = 
{\lim}_{N \rightarrow \infty} \frac{d}{d\ln N} \left( 
\ln <N|\eta |> \right) ~~~\ , 
\end{eqnarray}
one can see that due to (\ref{part}), the scaling (\ref{first}) holds when 
$g=(\gamma + \beta )/(\gamma + 2\beta )$~. Consequently, $g$~ is 
contained between 1/2 and 1. Whenever the cluster-size can be
reasonably defined for the second-order transition, 
like it is the case in percolation,
Ising model or Fisher droplet model, 
the exponent $\tau$~ of the power-law cluster-size
distribution : $n(k) \sim k^{-\tau }$~, satisfies additional
relations\cite{Stauffer} : 
$\gamma + \beta =1/\sigma$ and $\gamma + 2\beta = (\tau -1)/\sigma$ 
~, which yield : $g \equiv 1/(\tau -1)$~. This means that 
$\tau$ has to be contained between 2 and 3 when (\ref{first}) holds.

What may happen if the order parameter is not known 
exactly? To illustrate this point, let us consider 
a quantity : $m=N^{\kappa } - N \eta$~, where $\eta $~ is
the true order parameter and 
$\kappa $~ is larger than the anomalous dimension $g$~. 
For large $N$~, $|m|$ is of order $N^{\kappa}$~. 
Writing (\ref{first}) with $m$ instead of $\eta$, 
and taking into account : $P[\eta ] d\eta =P[m]dm$~, 
one finds the delta - scaling :
\begin{eqnarray}
\label{delta}
<|m|>^{\delta }P[m] = \Phi (z_{({\delta})}) 
\equiv \Phi \left( \frac{m-<|m|>}{<|m|>^{\delta}} 
\right) ~~\ , ~~~~~ 
\delta = \frac{g}{\kappa} < 1 ~~~\ ,
\end{eqnarray}
with the scaling function ${\Phi} (z_{(\delta )})$~ 
depending only on the scaled variable :
$z_{(\delta )} = (m-<|m|>)/<|m|>^{\delta}$~. According 
to (\ref{psi}) and (\ref{first}),
the logarithm of scaling function :
\begin{eqnarray}
\label{logar}  
\ln \Phi (z_{(\delta )}) = -{\beta}_T f_o(z_{(\delta )}, c) ~~~ \ ,
\end{eqnarray}
is directly related to the non-critical free energy 
, in either  ordered ($c>0$) or disordered $(c<0)$ 
phase. The relation $\delta = g/\kappa $~ in (\ref{delta}) 
singularizes importance of an
extensive variable : $m=N(1-\eta )\equiv N{\hat {\eta}}$~. ${\hat {\eta}}$~ 
can be useful in phenomenological applications and plays an important role in
the percolation studies\cite{Stauffer}~. One finds for this choice : 
$<m> \sim N$~, with algebraic finite-size corrections, and  the 
delta-scaling (\ref{delta}) with $\delta = g$. Hence, 
$P[N{\hat {\eta}}]$~ allows to determine the anomalous dimension $g$~ and,
consequently, the ratio of critical exponents $\beta$~ and $\gamma$~.

Let us suppose now that the extensive parameter $m $ is not critical 
, {\it i.e.}, either the system is in
a critical state but the parameter $m$ is not critical, 
or the system is
outside of critical region. The value of $m$ at the 
equilibrium is
obtained by minimizing the free energy. Let us suppose 
that the 
free energy $F$ is analytical in the variable $m$ close 
to its most probable
value $m^{*}$, {\it i.e.}, :
\begin{eqnarray}
\label{freenot}
F \sim N^{-\phi} (m-m^{*})^{{\phi}+1} ~~~ \ .
\end{eqnarray}
In most cases ${\phi}=1$~, though, in general, ${\phi}$~ can take
any positive integer value. Using (\ref{freenot})
one obtains : $<|m|> \sim \mu^{*} N$~, where $\mu^{*}$ is 
a positive (finite) number independent of $N$~, and : 
\begin{eqnarray}
\label{zkonc}
Z_N \sim N^{\frac{{\phi}}{{\phi}+1}} \sim 
~<|m|>^{\frac{{\phi}}{{\phi}+1}} ~~~ \ .
\end{eqnarray}
The probability density $P[m]$ verifies the 
generalized scaling law (\ref{delta}) :
\begin{eqnarray}
\label{second}
<|m|>^{\delta} P[m] = \Phi (z_{(\delta 
)}) =  
\exp \left( -{\beta}_T \mu^{*{\phi}}
\left( \frac{(m-<|m|>)}{<|m|>^{\frac{{\phi}}{{\phi}+1}}} 
\right)^{{\phi}+1} \right) ~~~ \ , 
\end{eqnarray}
but now $\delta $ ($= {\phi}/({\phi}+1) < 1$) 
is constrained by the value of ${\phi}$~.
In the generic case ${\phi}=1$~, $\delta $~ equals 1/2
and the scaling function is Gaussian\cite{comment2}. 
The second scaling (\ref{second}) holds
for $<m> \sim N$ but now with the exponential finite-size corrections.

\section{Results}\label{sec:results}
The above results apply to any {\it second} order transition,
and , in particular, they are not limited 
to the Landau-Ginzburg theory of phase transitions. 
\begin{figure}
\vspace{8cm}
\includegraphics{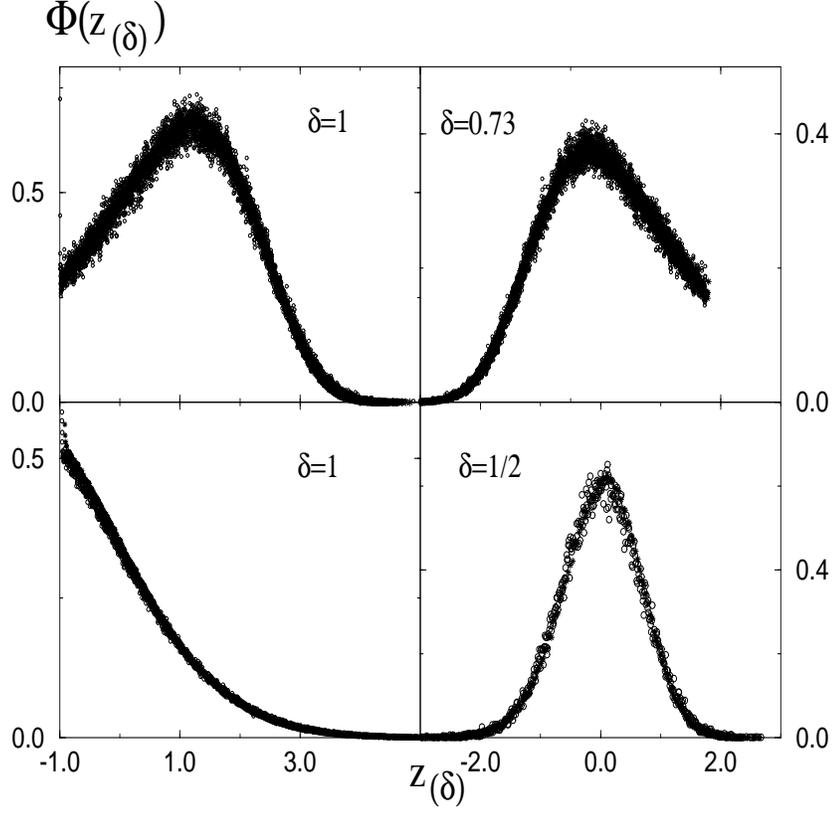}
\label{fig1}
\caption{ The order parameter fluctuations in the bond percolation model in 3D are
calculated in the lattices : $N = 10^3 $ and $N=14^3$~, and plotted in the
scaling variables of delta-scaling (\ref{delta}). {\bf (a)} 
(the upper left plot)
: for the bond activation probability : $p_{cr}=0.2482$~; {\bf (b)} (the upper
right plot) : for $p=0.245 \simeq p_{cr}$~; {\bf (c)}
(the lower left plot) : for $p=0.35 > p_{cr}$~; {\bf (d)}
(the lower right plot) : fluctuations 
of the quantity : $m = M_1 - S_{max}$~, 
where $M_1 = \sum_k kn(k)$~ is the first moment of 
the fragment-size probability distribution, at the percolation threshold 
$p_{cr}$ are plotted in the scaling variables of delta-scaling (\ref{delta}) 
with $\delta \simeq 0.80$~. For more details, see the description in the text.}
\end{figure}
As an illustration, let us look at the  
bond percolation model ( Fig. 1). The intensive order parameter
is the normalized mass of 
largest cluster : $\eta = S_{max}/N$~. Fig. 1a (the upper left plot) 
shows the scaling function $\Phi (z_{(1 )})$~ in $3D$-percolation at
the value of bond activation probability ($p_{cr}=0.2482$) 
corresponding to the critical activation in the infinite system. 
Results for different lattice sizes 
are superposing well in the scaled variables 
(\ref{delta}) for $\delta = 1$~. 
One should recall that ${\Phi}(z_{(1)})$~  
is also a 'portray' of the free energy density 
at a renormalized distance from the critical point. The
probability distribution $P[N\eta ]$~ changes rapidly, as can be 
seen in Figs. 1a and 1b (the upper right plot). 
The latter plot is obtained for the bond activation probablity 
$p=0.245$~, very close to $p_{cr}$. 
Results in Fig. 1b are plotted in the scaling
variables (\ref{delta}) for $\delta = 1$~, even though 
slight deviations between different calculations can be seen 
in the shoulder region and in the tail for positive $z_{(1)}$~. Fig.
1c (the lower left plot) shows the order parameter 
distribution in the 'liquid phase' ($p=0.35 > p_{cr}$)~.  One finds  
the second scaling (\ref{second})~, 
in agreement with the analytical derivation. Finally, Fig. 1d
(the lower right plot) shows
what happens at the percolation threshold ($p = p_{cr}$) 
if instead of $P[N\eta ]$~, one plots 
the probability distribution of $m = M_1 - S_{max}$~, 
where $M_1 = \sum_k kn(k)$~ is the first moment of 
the fragment-size probability distribution.
$m$~ ($\equiv N{\hat {\eta}}$~), 
 is related in a non-trivial way to the order
parameter and, in particular, it conserves 
the singularity of $S_{max}$~. $P[m]$~ obeys the delta-scaling
(\ref{delta}) for $\delta \simeq 0.80$~, what should be compared with 
: $1/(\tau -1) = 0.84$~, given by the analytical argument leading to 
the delta-scaling (\ref{delta})
. This signature of phase transition 
disappears for $p \neq p_{cr}$~, {\it i.e.}, $\delta $~ becomes equal 1/2.

\begin{figure}
\vspace{8cm}
\includegraphics{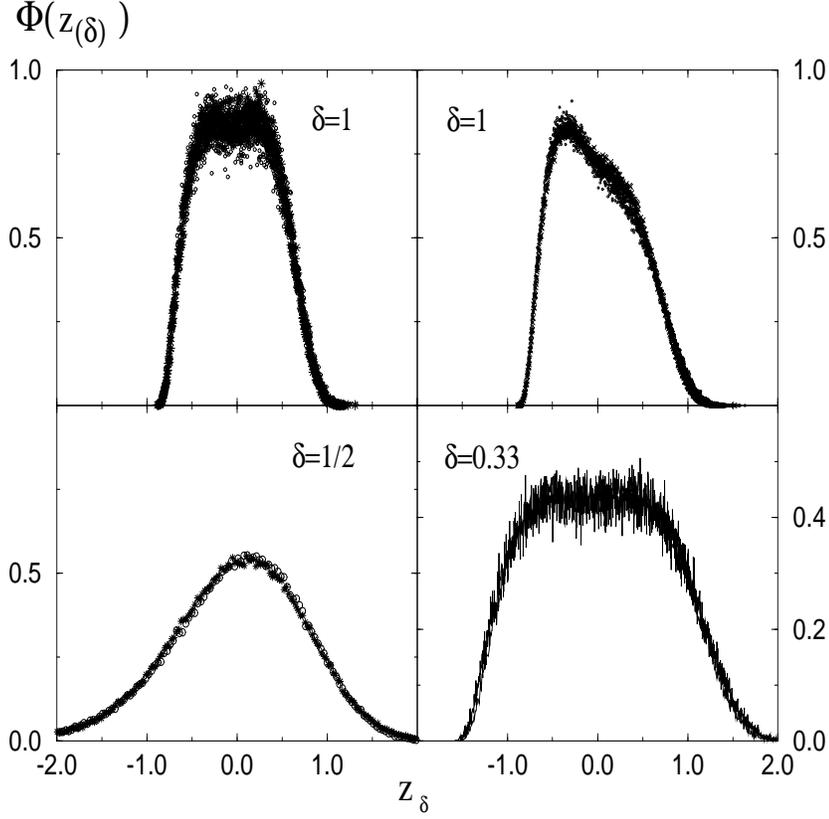}
\label{fig2}
\caption{The order parameter fluctuations in the FIB model calculated 
for two initial sizes : $N=2^{10}$ and $N=2^{14}$~, and plotted in the
scaling variables of delta-scaling (\ref{delta}). 
{\bf (a)} (the upper left plot) : the critical FIB process for 
$p_F=0.875$~, $a=0$~ corresponding to the anomalous dimension $g=0.75$~; 
{\bf (b)} (the upper right plot) : fluctuations of the quantity :
$m = N - M_0$~ are plotted in the scaling variables 
of delta-scaling (\ref{delta}) 
with $\delta \simeq 0.73$~, {\bf (c)} (the lower left plot) : 
the critical FIB process for 
$p_F=0.7$~, $a=0$~ corresponding to the anomalous dimension $g=0.4$~;
{\bf (d)} (the lower right plot) : shattered phase for $a=0,~ b=-1,~ \tau =4$~.
For more details, see the description in the text.}                         
\end{figure}
As a second example, let us consider the FIB
model \cite{sing1,sing11}~ which exhibits the second order, shattering phase 
transition \cite{Ziff}. In this
off-equilibrium case, analytical derivation of (\ref{first}) and
(\ref{delta}), in principle, does not apply. One deals in FIB model  
with clusters characterized by a conserved 
quantity, called the cluster mass. The anscestor cluster of mass $N$~
is fragmenting via an ordered and irreversible sequence 
of steps until either the cutoff-scale for monomers
is reached or all clusters are inactive.

Each step in this cascade is either a binary 
fragmentation of an active cluster
$(k) \rightarrow (j)~+~(k-j)$~ with a fragmentation rate
$\sim F_{j,k-j}$~ (a mass partition probability), or its
inactivation $(k) \rightarrow (k)^{*}$~ with an
inactivation rate $\sim I_k$~.
The order parameter is here the 
reduced cluster multiplicity $M_0/N$~ or 
the reduced monomer multiplicity, both of them closely 
interrelated.                                  
The total fragmentation probability $p_F$~
at each step of FIB cascade is :
\begin{eqnarray}
\label{pfdef}
p_F(k) = \sum_{j=1}^{k-1} F_{j,k-j}~( I_k +
\sum_{j=1}^{k-1} F_{j,k-j} )^{-1} ~~~ \ .
\end{eqnarray} 
The cluster mass 
independence of $p_F(k)$~ at any step until the 
cutoff-scale characterizes the {\it critical transition 
region}. FIB process is self-similar in this regime. 
For $p_F > 1/2$~, the anomalous dimension (\ref{anomdim}) 
(now $N| \eta | \equiv M_0$)~
varies from 0 to 1, what is different from the limits on $g$~ in the 
equilibrium systems. The average multiplicity of inacitve 
clusters is \cite{kno}~: 
$<M_0> \sim N^{\tau - 1}$~ ($1 < \tau < 2)$~, and $g=\tau -1$~. 
In the {\it shattered phase}, the average multiplicity 
is : $<M_0>~ \sim N$~, 
the cluster-size distribution is a
power-law with $\tau > 2$~ and the anomalous dimension is
$g = 1$~. In this phase, $p_F$~ is an increasing  
function  of cluster mass $k$~ and, hence, the FIB cascade is not self-similar.
 Most of the interesting physical applications correspond to
homogeneous fragmentation functions \cite{sing11} : $F_{\lambda j, 
\lambda (N-j)} = {\lambda }^{2a} F_{j,N-j}$~, {\it e.g.},  
$F_{j,N-j} \sim [j(N-j)]^{a}$. 
For the homogeneous inactivation rate-function : $I_k 
\sim k^{b}$, the critical
transition region in FIB model corresponds to the line : 
$b = 2a + 1$~, and the shattered phase
corresponds to : $b < 2a + 1$~. Such homogeneous
rates : $F_{j,k-j}$~ and $I_k$~
, will be used in examples shown in Fig. 2.
Fig. 2a (the upper left plot) 
shows the scaling function $\Phi (z_{(1 )})$~ of  
critical FIB process for 
$p_F=0.875$~, $a=0$~, what yields $g=0.75$~. 
The asymmetry of $\Phi (z_{(1)})$~ about $z_{(1)}=0$~, is rather common
in the critical sector of FIB model \cite{kno}. 
This sector and its characteristic first
scaling (\ref{first}) extends to the domain $0 < g < 1/2$~\cite{kno}~,
excluded in equilibrium models. In this domain ($g<1/2$)~, 
the most probable value of ${\Phi}(z_{(1)})$ is at $z_{(1)}=-1$~, 
whereas for $g>1/2$~ it takes a 
value close to $z_{(1)}=0$~\cite{kno}~. This can also be seen in 
Fig. 2c (the lower left plot) which exhibits the 
scaling function $\Phi (z_{(1 )})$~ of critical FIB process for
$p_F=0.7$~, $a=0$~, for which $g=0.4$~. What happens 
if instead of $P[M_0] \equiv P[N\eta ]$~, one 
plots $P[N-M_0] \equiv P[N{\hat {\eta}}]$~, 
as shown in Fig. 2b (the upper right plot). 
Analogously as in percolation (see Fig. 1b), $P[N{\hat {\eta}}]$~ 
is scaling (\ref{delta}) with the non-trivial exponent $\delta \simeq 
0.73$~ which is close to the value $\delta = g = 0.75$~, obtained using
analytical arguments (\ref{delta}). 
The order parameter distribution in the
shattered phase ($a=0, b=-1, \tau =4$)~, 
is shown in Fig. 2d (the lower right plot) in the 
variables of second scaling (\ref{second})~ for $\delta = 1/2$~.
Again, one finds an analogy to the situation  
in the 'liquid' phase of percolation (Fig. 1c).

\section{Outlook}
The off-equilibrium FIB model shows essentially the same
relation between criticality and scaling of 
order parameter fluctuations as it has been derived analytically for the 
second-order equilibrium
transitions (Eqs. (\ref{first}), (\ref{delta}), (\ref{second})).
This is related to the
underlying self-similarity which is common to both equilibrium and 
off-equilibrium realizations of the second-order transition.  In that
sense, the scaling laws (Eqs. (\ref{first}), (\ref{delta}), (\ref{second})) are
the salient features of {\it any} system exhibiting the
second-order transition and the function ${\Phi}(z_{(\delta )})$~ is a
fingerprint of the system and its transition. It is
therefore important to study implications of 
scaling (\ref{first}) on statistical properties of the system.
Let us first postulate the definition of the pseudo-free energy :
\begin{eqnarray}
\label{pseudo}
{\cal F} = - {{\tilde {\beta}_T}}^{-1} \ln(<| \eta |>P[ \eta ]) ~~~ \ ,
\end{eqnarray}
with a coefficient ${\tilde {\beta}_T}$ which is independent of $\eta $~ and 
characterizes the homogeneous system.        
If we suppose that the first scaling (\ref{first}) holds and employing the
asymptotical definition
of the anomalous dimension (\ref{anomdim})~, one finds :
\begin{eqnarray}
\label{pseudoscal}
{\cal F}/N \sim {\eta}^{1/(1-g)} \phi (\eta N^{1-g}) ~~~ \ ,
\end{eqnarray}                       
what formally corresponds to (\ref{Widomfinite}) with $(1-g)$ instead of 
both $\beta/(2-\alpha ) $~ and 
$\beta /(\nu \delta )$. ${\cal F}$ appears to be a pseudo-free 
energy for {\it constrained}
$\eta $~, and {\it fixed} both $N$ and ${\tilde {\beta}_T}$.  
This derivation does not suppose that the system is
close to the second-order transition in
thermodynamical equilibrium. In particular, (\ref{first}) and 
(\ref{anomdim}) are expected to hold for any critical
behaviour : at the thermodynamic equilibrium as we have shown rigorously, but
also for the non-thermodynamic equilibrium like in percolation, for the
off-equilibrium final state like in FIB model, or for the
self-organized criticality \cite{SOC}. The basic quantity is ${\cal F}$~,
which contains thermodynamically relevant information about the 
analogue of the inverse of temperature. 
The scaling law of ${\cal F}$~ ~(\ref{pseudoscal})
gives in turn information about the critical behavior of the system.
Finally, let us remind that the formulae (\ref{pseudoscal}) contains only one
exponent $g$. To have an access to other
critical exponent, we need to vary the field conjugate to
the order parameter.

To summarize, we have found the new 
approach to study critical phenomena both in
equilibrium and off-equilibrium systems. This approach is based on the
existence of universal scaling laws in the probability distribution of both the 
order parameter and its complement, in the second order phase transitions. The
precise relation between the scaling functions ${\Phi}(z_{(\delta)})$~ 
, the nature of order parameter and the critical exponents
yields a new tool for determining the combinations of critical exponents
even in small systems and for learning about the nature of critical phenomenon.
We hope, this approach will be useful in many phenomenological applications in
the strong interaction physics and in the condensed matter physics.


\section*{References}
\begin{thebibliography}{99}
\bibitem{Widom}
B. Widom, \Journal{\JCP}{43}{3898}{1965};\\
F.J. Wegner, {\em Phase Transitions and Critical 
Phenomena}, vol.6, pps. 8-122,
C. Domb and M.S.Green eds, Academic Press (London, 
1976).

\bibitem{FisherBarber}
M.E. Fisher and N.N. Barber, \Journal{\PRL}{28}{1516}{1972}.

\bibitem{FF}
M.E. Fisher and A.E. Ferdinand, \Journal{\PRL}{19}{169}{1967}.

\bibitem{comment}
The constant $c$~ is negative if the maximum of the 
finite-size susceptibility or of any 
other divergent macroscopic quantity in the 
thermodynamic limit lies in
the ordered phase, and is positive if this maximum is in 
the disordered phase.


\bibitem{Mayer}
J.E. Mayer and M.G. Mayer, {\em Statistical Mechanics}, 
John Wiley and Sons (London, 1957).


\bibitem{sing1}
R. Botet and M. P{\l}oszajczak, \Journal{\PRL}{69}{3696}{1992}.



\bibitem{sing11}
R. Botet and M. P{\l}oszajczak, \Journal{\IJMP}{3}{1033}{1994}.


\bibitem{koba} 
Z. Koba, H.B. Nielsen, and P. Olesen, \Journal{\NPB}{40}{317}{1972}.

 
\bibitem{comment1}
This follows from expected validity of Feynman scaling 
for 
the many-body inclusive cross sections of particle
production in ultrarelativistic collisions.


\bibitem{antoniou}
N.G. Antoniou, A.I. Karanikas, and S.D.P. Vlassopoulos, 
\Journal{\PRD}{29}{1470}{1984}; 
\Journal{\PRD}{14}{3578}{1976}.



\bibitem{kno}
R. Botet and M. P{\l}oszajczak, \Journal{\PRE}{54}{3320}{1996}.

\bibitem{Stauffer}
D. Stauffer and A. Aharony, {\em Percolation Theory}, (second edition) 
Taylor and Francis (London, Washington DC, 1992).


\bibitem{Ziff}
E.D. McGrady and R.M. Ziff, \Journal{\PRL}{58}{892}{1987};\\
Z. Cheng and S. Redner, \Journal{\PRL}{60}{2450}{1988}.


\bibitem{comment2}
In general, the second scaling law
(\ref{second}) can be satisfied in 
a broader class of scaling functions than Gaussian functions\cite{prln}~. 


\bibitem{prln}
R. Botet,  M. P{\l}oszajczak, and V. Latora, \Journal{\PRL}{78}{4593}{1997}.


\bibitem{SOC}
P. Bak, C. Tang and K. Wiesenfeld, \Journal{\PRL}{59}{381}{1987};\\
C. Tang and P. Bak, \Journal{\PRL}{60}{2347}{1988};\\
P. Bak, K. Chen and C. Tang, \Journal{\PLA}{147}{297}{1990}.



\end{thebibliography}
\end{document}